\documentclass[prl,aps,twocolumn,superscriptaddress,showkeys,floatfix]{revtex4}

\usepackage{graphicx}
\usepackage[latin1]{inputenc}
\usepackage{hyperref}

\begin{document}

\title{Imaging Spin Reorientation Transitions in Consecutive Atomic Co layers on Ru(0001)}

\author{Farid El Gabaly}
\affiliation{Dpto. de F\'{i}sica de la Materia Condensada and Centro de Microan\'{a}lisis de Materiales, Universidad Aut\'{o}noma de Madrid, Madrid 28049, Spain}
\email{farid.elgabaly@uam.es}
\homepage[]{http://hobbes.fmc.uam.es/loma}
\author{Silvia Gallego}
\author{Carmen Mu\~{n}oz}
\affiliation{Instituto de Ciencia de Materiales de Madrid, CSIC, Madrid 28049, Spain}
\author{Laszlo Szunyogh}
\affiliation{Department of Theoretical Physics and Center for Applied Mathematics and Computational Physics, Budapest University of Technology and Economics, Budafoki \'ut 8, H-1521 Budapest, Hungary}
\author{Peter Weinberger}
\affiliation{Center for Computational Materials Science, Vienna University of Technology, Gumpendorferstr. 1a, A-1060 Vienna, Austria}
\author{Christof Klein}
\author{Andreas K. Schmid}
\affiliation{Lawrence Berkeley National Laboratories, Berkeley, California 94720, USA}
\author{Kevin F. McCarty}
\affiliation{Sandia National Laboratories, Livermore, California 94550, USA}
\author{Juan de la Figuera}
\affiliation{Dpto. de F\'{i}sica de la Materia Condensada and Centro de Microan\'{a}lisis de Materiales, Universidad Aut\'{o}noma de Madrid, Madrid 28049, Spain}
\email{juan.delafiguera@uam.es}

\begin{abstract}
By means of spin-polarized low-energy electron microscopy (SPLEEM), we show that the magnetic easy-axis of one to three atomic-layer thick cobalt films on Ru(0001) changes its orientation twice during deposition: one-monolayer and three-monolayer thick films are magnetized in-plane, while two-monolayer films are magnetized out-of-plane. The  Curie temperatures of films thicker than one monolayer are well above room temperature. Fully-relativistic calculations based on the Screened Korringa-Kohn-Rostoker (SKKR) method demonstrate that only for two-monolayer cobalt films the interplay between strain, surface and interface effects leads to perpendicular magnetization.
\end{abstract}

\date{\today}


\maketitle 
Applications of ferromagnetic films depend on understanding and controlling the direction of the easy-axis of magnetization. In particular, magnetization perpendicular to the film plane\cite{Gradmann74,Carcia85,Engel91,Johnson1996} holds promise for novel information-processing technologies\cite{Chappert}. Two important features of ultra-thin films underlie this technological achievement: the high Curie temperature of transition metal films and the ability to control their microstructure.
To provide deeper understanding, we study thin-film magnetism in a system whose components do not intermix, Co and Ru. Previous work\cite{Panissod1992PRB,Dennis2002} has shown that the easy axis of magnetization in Co/Ru multilayers changes from perpendicular at low Co thickness to in-plane for films thicker than 7 ML\cite{Muller1992}.  Because the Co films did not grow layer-by-layer\cite{Bader1993, Ding2005}, the films contained islands of varying thickness.  Under these conditions, determining precisely how the magnetization changes  as a function of film thickness is quite problematic.

\begin{figure}
\centerline{\includegraphics[width=0.5\textwidth]{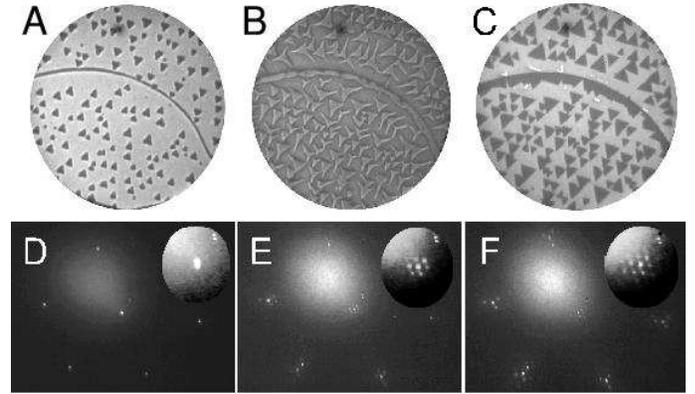}}
\caption{\label{triang}LEEM images and diffraction patterns of a Co film growing on Ru(0001). (a)-(c): LEEM images show the morphology of the growing film. Field of view is 10~$\mu$m, electron energy is 5 eV, and growth temperature is 460~K. One single, curved Ru step crosses the images.  (a) 1~ML Co islands (dark) on Ru (light gray background). (b) 2~ML islands (light gray) on a complete 1~ML film (dark gray). (c) 3~ML islands (dark gray) on a nearly complete 2~ML film (light gray). (d)-(f)  LEED patterns (70 eV) obtained from selected film areas of uniform thickness. Insets show magnified views of the specular beam. (d) 1~ML , (e) 2~ML and (f) 3~ML of Co/Ru(0001).}
\end{figure}

Here we deposit Co films under conditions of perfect layer-by-layer growth. Then we use in-situ spin-polarized low-energy electron microscopy (SPLEEM)\cite{Bauer1994,Duden1995,Duden1998} to locally determine the magnetization orientation of one-, two-, and three-monolayer thick Co films. We observe that the easy axis of magnetization changes after the completion of each atomic layer.  By combining structural, morphological and microscopic magnetic measurements with fully relativistic ab-initio calculations based on the screened Korringa-Kohn-Rostoker (SKKR)\cite{Weinberger2005} method, we explain the origin of the magnetization changes.
Our results highlight that the magnetic anisotropy of ultra-thin films is not simply explained by strain or interface effects alone, but often by a combination of both effects.

The films are grown in two different ultra-high vacuum low-energy electron microscopes (LEEM and SPLEEM)\cite{Bauer1994} by physical vapor deposition from calibrated dosers at rates of 0.3~ML/min. Details of the substrate-cleaning procedure as well as the experimental system are given elsewhere\cite{Farid2005}. Perfect layer-by-layer Co growth occurs up to at least 7~ML when the Ru substrate has a low density of atomic steps (Fig.\ref{triang}a-c). Because substrate steps enable a kinetic pathway to the nucleation of new film layers, three-dimensional growth\cite{Muller1992,Bader1993} occurs after the first monolayer if substrate steps are present at even moderate density\cite{Ling2004a}. The film structure is determined by selected-area low-energy electron diffraction (LEED), i.e., the diffraction patterns were acquired with diffracted electrons coming from areas of the film with uniform thickness. One-monolayer films always present a 1$\times$1 LEED pattern indicating pseudomorphic growth, that is,  the film has the same in-plane lattice parameter as the substrate (Fig.~\ref{triang}d). Since the in-plane lattice parameter of bulk Co is 7.9~\% smaller than that of Ru, both measured within the hexagonal-close-packed (hcp) basal plane, the first monolayer of Co is under pronounced tensile strain. Analysis of the intensity versus energy curves of the specular and integer diffraction spots establishes that the Co film continues the hcp stacking\cite{hwang92} of the substrate, with a Co-Ru interplanar separation estimated to be contracted ~6\% relative to the Ru-Ru interplanar spacing.
For films thicker than 1~ML, satellite spots appear around the bulk diffraction beams (Fig.~\ref{triang}e-f), i.e., the thicker films are no longer pseudomorphic. From the diffraction patterns, we estimate that the in-plane spacing of 2~ML and 3~ML Co films is 5$\pm$1~\% less than the Ru spacing, leaving the film strained only by 3~\% relative to the bulk-Co value.
At intermediate coverages between 1~ML and 2~ML, the 1~ML areas are still pseudomorphic, as detected by dark-field imaging\cite{Farid2005}, while 2~ML islands are relaxed and 3~ML films grow mainly in a face-centered-cubic structure.

\begin{figure}
\centerline{\includegraphics[width=0.5\textwidth]{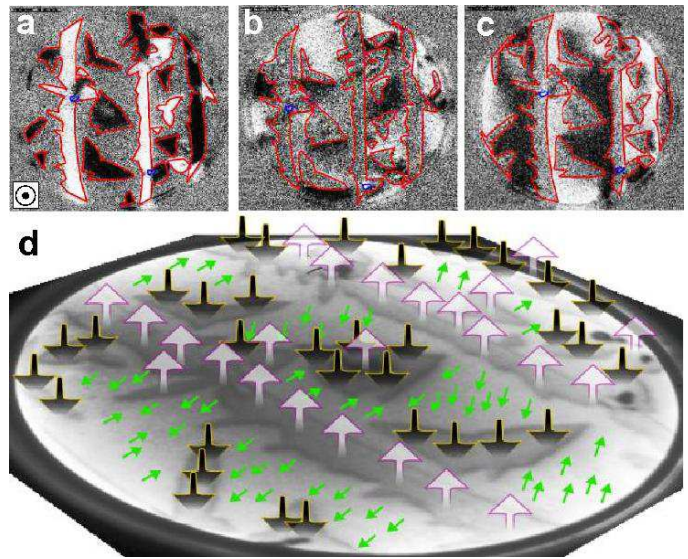}}
\caption{\label{exp1}(color) Images of topography and magnetization of one region of a 1.5 ML Co/Ru(0001) film. Images were taken at 110~K. Field of view is 2.8~$\mu$m and electron energy is 7 eV. (a)-(c) SPLEEM images with electron-polarization oriented: (a) out-of-plane; (b) in-plane and 13$^\circ$ off a compact-direction; (c) in-plane and 103$^\circ$ off a compact-direction. 2~ML islands are framed in red (two small 3~ML islands are framed in blue). (d) LEEM image of the surface with the deduced magnetization direction indicated by arrows (black and white arrows mean out-of-plane magnetization, green arrows mean in-plane magnetization). Dark gray indicates 2~ML islands, light gray 1~ML film.}
\end{figure}

\begin{figure}
\centerline{\includegraphics[width=0.5\textwidth]{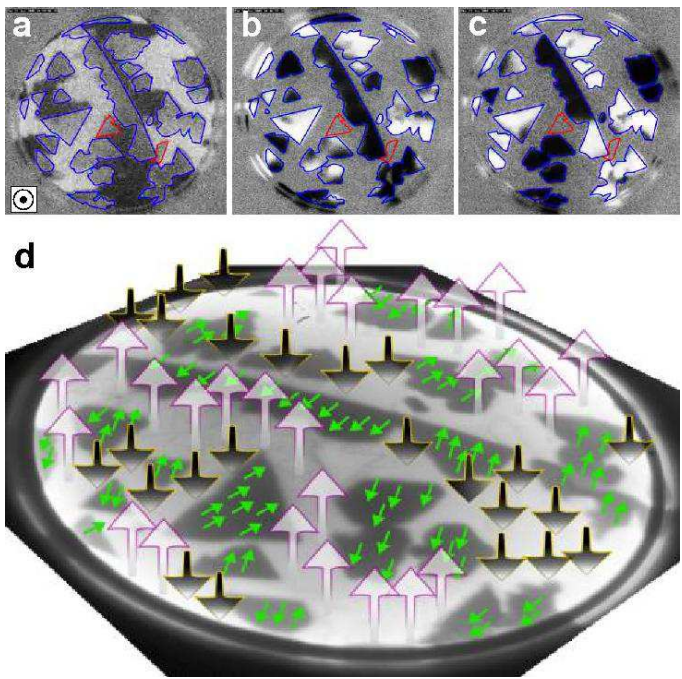}}
\caption{\label{exp2}(color) Images of topography and magnetization of one region of a 2.5~ML Co/Ru(0001) film. Images were taken at room temperature. Field of view is 2.8~$\mu$m and electron energy is 7 eV. (a)-(c) SPLEEM images with electron-polarization oriented: (a) out-of-plane; (b) in-plane and 13$^\circ$ off a compact-direction; (c) in-plane and 103$^\circ$ off a compact-direction. 3~ML islands are framed in blue. Two vacancy-islands in the 2~ML area, where Co is 1~ML thick, are framed in red. (d) LEEM image of the surface with the deduced magnetization direction indicated by arrows (black and white arrows mean out-of-plane magnetization, green arrows mean in-plane magnetization). Dark gray indicates 3~ML islands, light gray 2~ML film.}
\end{figure}

To characterize the easy axis of magnetization, we employ SPLEEM\cite{Duden1998}. With this technique the magnetization can be mapped onto three orthogonal directions~\cite{Ramchal2004PRB}:
the absence of contrast in the images (gray) indicates no magnetization component along the selected direction; bright and dark areas indicate a component of the magnetization along or opposed to the illuminating beam polarization, respectively.
In Fig.~\ref{exp1} we show LEEM and SPLEEM images of a film that consists of a complete monolayer of Co plus some second layer islands (Fig.~\ref{exp1}a), both in the middle of the substrate terraces and at the bottom of the ruthenium substrate steps. The SPLEEM images show the spatially resolved component of the magnetization in three orthogonal directions: two in-plane (Fig.~\ref{exp1}b-c) and one out-of-plane (Fig.~\ref{exp1}d). In one-monolayer areas the magnetization is oriented in the plane of the film, while for two layer islands the magnetization is out-of-plane. For a complete 2~ML film with additional 3~ML islands (Fig.~\ref{exp2}), the magnetization of the 2~ML areas is out-of-plane. In contrast, 3~ML thick islands and thicker films (not shown) are magnetized in-plane.  To summarize, two magnetization easy-axis reorientation transitions are found in three consecutive atomic layers: at the crossover between 1 and 2~ML, and between 2 and 3~ML. This behavior has also been confirmed in films devoid of islands. We do not find intermediate easy-axis orientations (i.e., in between in-plane and out-of-plane), such as observed for Co films on other substrates\cite{Fritzsche1995,Duden1996}. The Curie temperature of the films changes dramatically from the first layer to the second. The first layer has a Curie temperature close to 170~K, as detected by the loss of magnetic contrast in the 1~ML areas. The Curie temperature of the 2~ML islands, which are magnetized out-of-plane, is well above room temperature, about 470~K. Thicker films exhibit Curie temperatures above 470~K. Iron films on W(110)\cite{Gradmann1999,Pietzsch2001} also present a double spin reorientation transition, but with a Curie temperature well below room temperature for out-of-plane magnetization\cite{Bergmann2004}. In this particular system, strain did not drive the reorientation transitions\cite{Bergmann2004,Prokop2005}.


The anisotropy energy that governs the orientation of the easy-axis of 
magnetization is the result of a delicate balance between different 
contributions. In thin films the dominating term is often the dipolar 
or shape anisotropy. This contribution, which results from the long-range magnetic dipole-dipole interactions, favors an in-plane orientation of the magnetization. 
However, other contributions such as the bulk, interface and surface 
magneto-crystalline anisotropy energies, as well as magneto-elastic 
terms\cite{Sander1999,Sander2004}, can compete with the dipolar anisotropy 
energy and can favor out-of-plane magnetization.
To understand the effects that give rise to the observed changes 
in the orientation of the Co magnetization, we perform ab-initio calculations 
in terms of the SKKR method\cite{Weinberger2005}. Changing the lattice 
parameters in the calculations allows us to determine how strain influences 
the magnetic anisotropy.
The magnetic anisotropy energy (MAE) is calculated as the difference of 
the total energy for in-plane and out-of-plane magnetization. 
A positive MAE corresponds to out-of-plane magnetization. 
By employing the force theorem\cite{Jansen99}, the MAE is defined as the sum 
of a band energy, $\Delta$E$_b$, and a magnetic dipole-dipole energy term, 
$\Delta$E$_{dd}$. The band-energy term can be further resolved into
contributions with respect to atomic layers that enable us to define surface and interface anisotropies. 

\begin{figure}
\centerline{\includegraphics[width=0.25\textwidth]{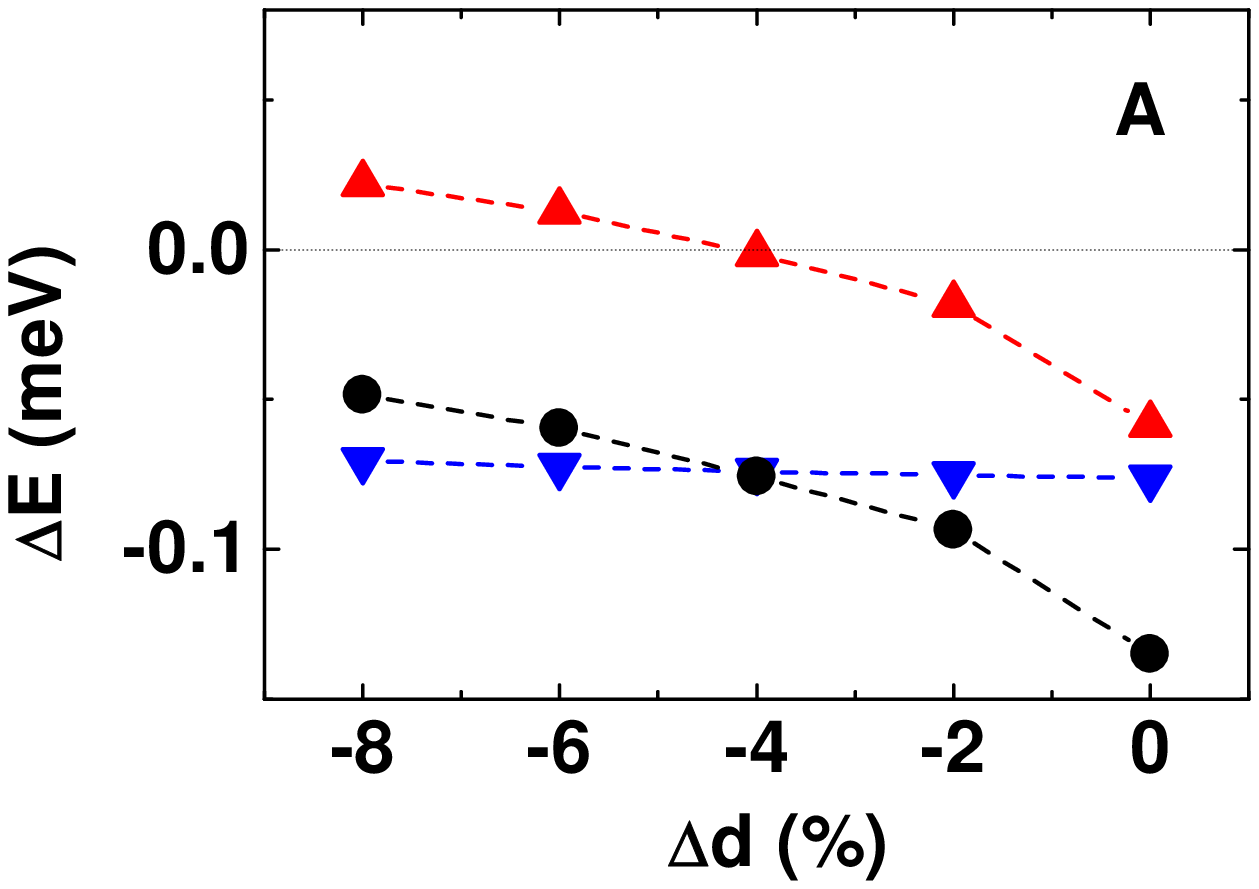}
\includegraphics[width=0.25\textwidth]{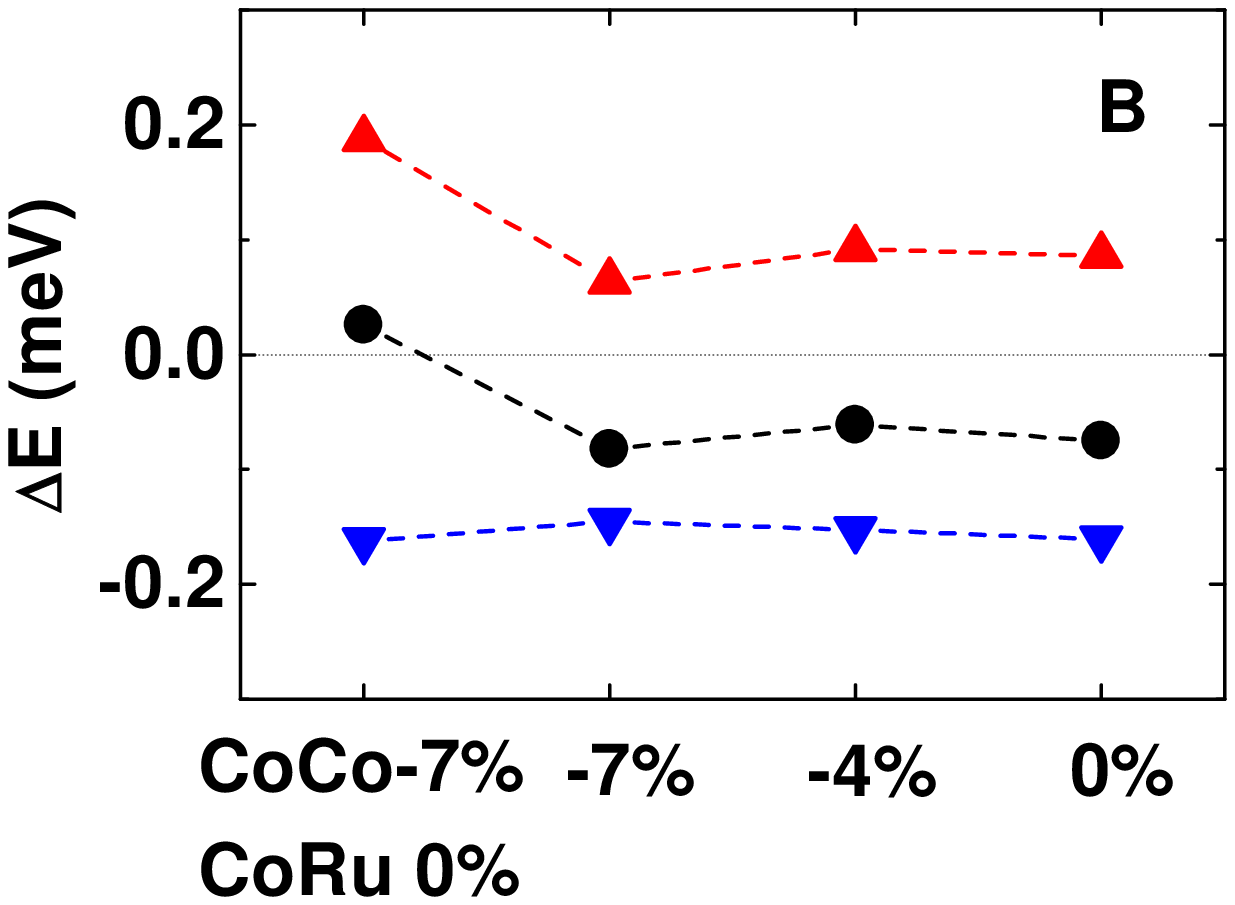}}
\centerline{\includegraphics[width=0.25\textwidth]{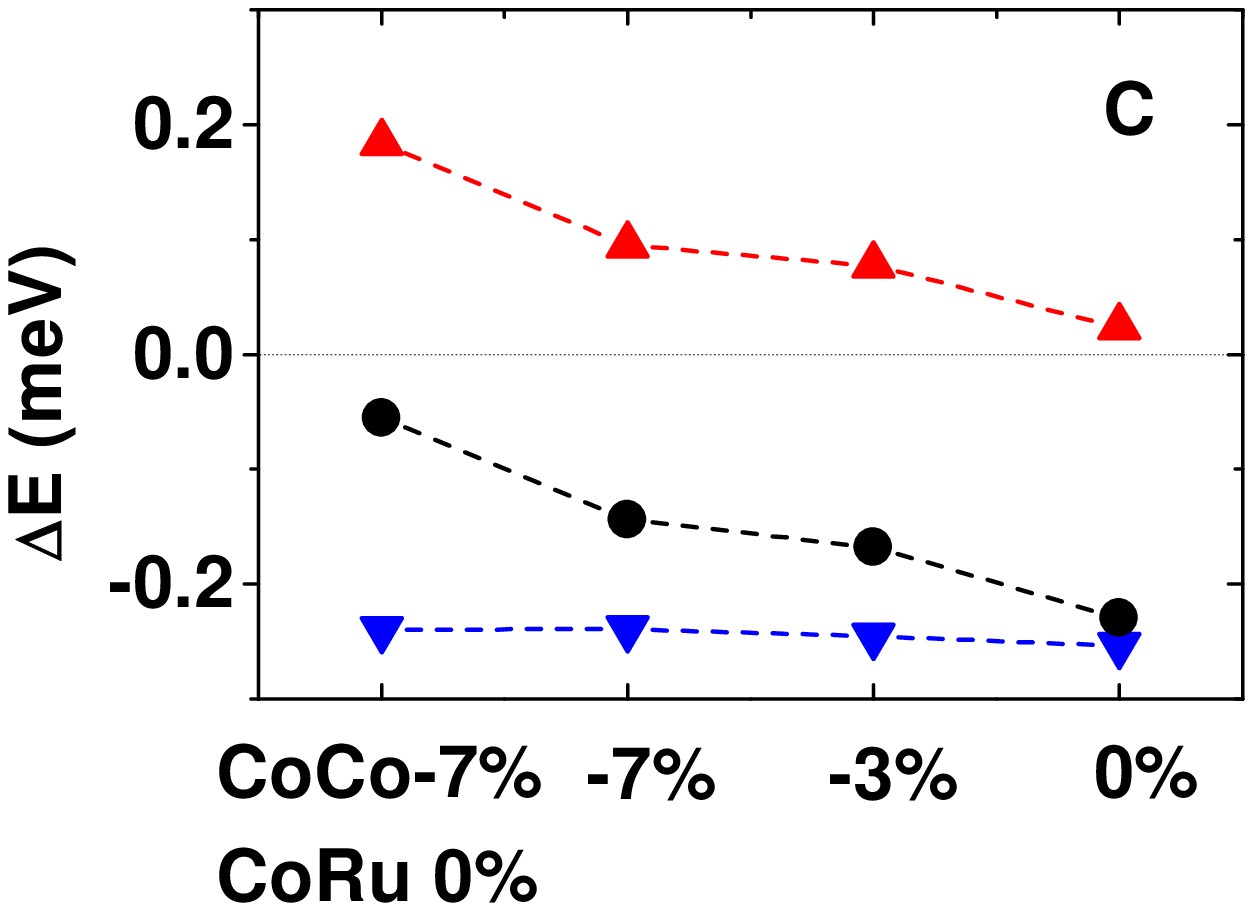}
\includegraphics[width=0.25\textwidth]{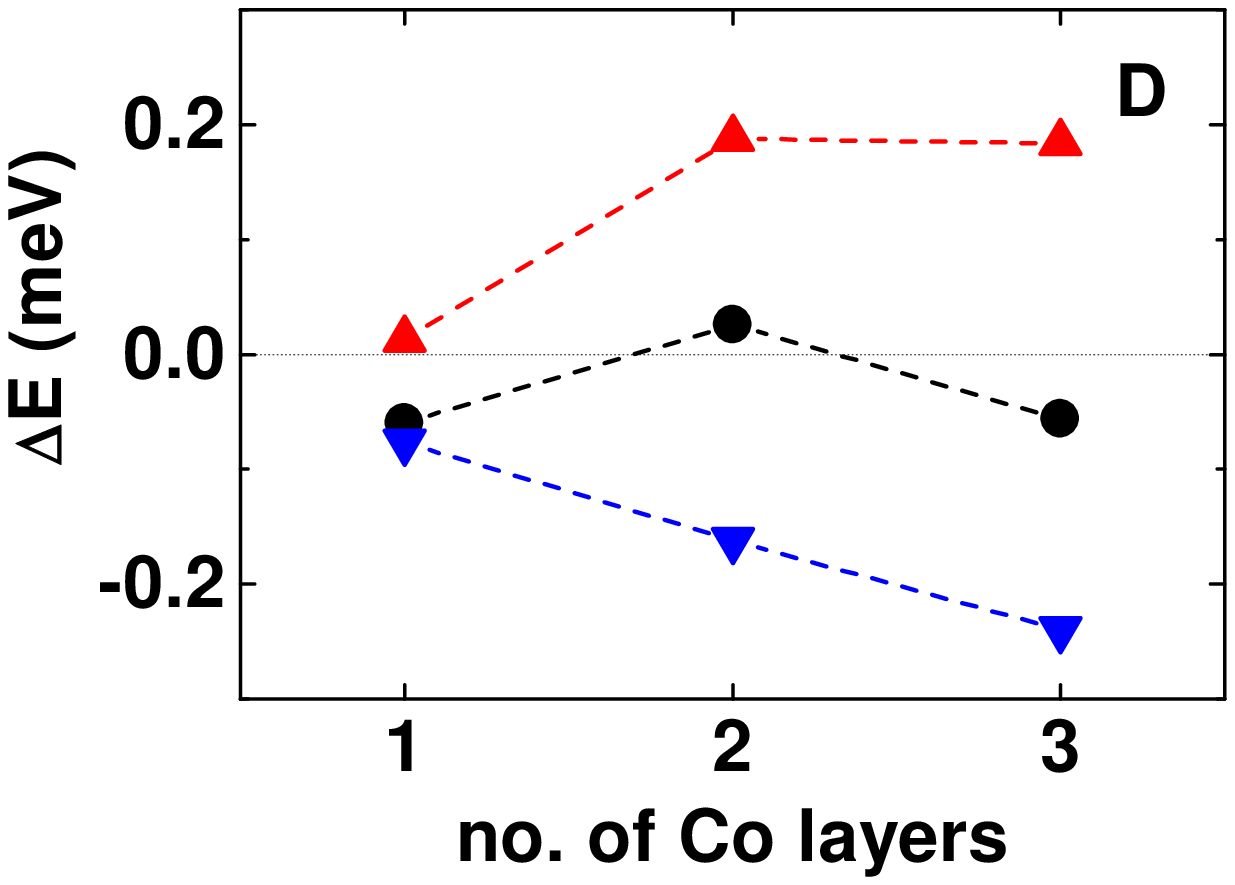}}
\caption{\label{calc}(color) Calculated magnetic anisotropy energies of the different Co films on Ru. (a)-(c) Dependence of the calculated MAE on the interlayer distance referred to the substrate interlayer spacing. MAE (black circle) and its components, $\Delta$E$_b$ (red up-triangle) and $\Delta$E$_{dd}$ (blue down-triangle) for: (a) a pseudomorphic 1~ML Co/Ru(0001) film under different contractions of the Co-Ru interlayer distance; in-plane strained (b) 2~ML Co/Ru(0001) and (c) 3~ML Co/Ru(0001) films with either the same or different (data points labelled by CoCo -7\% and CoRu 0\%) Co-Co and Co-Ru interlayer separations. (d) MAE and its components in the most realistic geometry for the 1, 2 and 3~ML Co films on Ru(0001), displaying the
double reorientation transition.}
\end{figure}

First, we calculate the anisotropy of the pseudomorphic one-monolayer Co films, taking into account contractions of the Co-Ru interlayer distance ($d$). As summarized in Fig.~\ref{calc}a, the value of $\Delta$E$_b$ increases as the interplanar spacing decreases; however, due to the negative $\Delta$E$_{dd}$, the preferred orientation of the magnetization remains always in-plane. Interestingly, the change in MAE is not proportional to the strain and, therefore, simple magnetoelastic arguments do not apply. Furthermore, we also tested the effect of contracting the in-plane lattice parameter of substrate and film. In that case, the MAE does not change significantly (result not shown in the figure). We conclude that the magnetization of the monolayer remains in-plane regardless of strain.

For two-monolayer and thicker films, the in-plane separation of the Co atoms is contracted by $\sim 5$\% with respect to the Ru structure, leading to a 20x20 coincidence lattice. We model the in-plane relaxation by contracting the supporting Ru substrate together with the Co film. Under this assumption, taking the same contraction for the Co-Co and Co-Ru interlayer spacing $d$ from 0 to 7\% relative to the substrate interlayer distance leads to a positive value of $\Delta$E$_b$ (Fig.\ref{calc}b) that, however, does not compensate the negative $\Delta$E$_{dd}$. For the bilayer, the observed positive sign of the MAE occurs  when different values for the Co-Co and Co-Ru interlayer distances are considered. In order to estimate the preferred relaxation of the interlayer distances we assume that atoms try to maintain the nearest-neighbors (NN) distances of their bulk materials, with Co-Ru distances being an average of the preferred Co-Co and Ru-Ru interlayer distances. This leads to contractions of 7\% for the Co-Co interlayer distances and a nearly unrelaxed Co-Ru spacing. As shown in Fig.~\ref{calc}b (leftmost data points), such a lattice distortion considerably increases $\Delta$E$_{b}$ resulting in a total positive MAE. A positive MAE is also obtained for an ideal Ru lattice with Co interlayer distances contracted by more than 4\% (not shown). In 3~ML thick films, non-uniform contractions of the Co layers lead also to an enhancement of the positive $\Delta$E$_b$ (Fig.\ref{calc}c). Nevertheless, the decrease in the $\Delta$E$_{dd}$ term associated with thicker films drives the magnetization in-plane. A summary of our calculations of the MAE for the Co films of different thickness, each at the most likely geometry, is shown in Fig.~\ref{calc}d. As a function of thickness the MAE changes sign twice, as observed experimentally.

Our calculations show that the double spin-reorientation transition is the result of a complicated interplay of structural and interface/surface electronic effects. All contributions to $\Delta$E$_b$ are strongly influenced by structural modifications.
For 2~ML thick films with the same Co-Co and Co-Ru interlayer separation, 
the dominant term is $\Delta$E$_b$ related to the interface Co.
However, when Co-Co and Co-Ru separations are allowed to be different, the contribution of the surface Co layer is remarkably enhanced resulting in a positive value of the MAE (out-of-plane magnetization).

In conclusion, we deposited films of Co onto Ru(0001) in the thickness range of up to 3 atomic monolayers and find that the Curie temperature is well above room temperature, provided the thickness is more than a single atomic monolayer. We observe two sharp reorientation transitions of the magnetization: 1~ML as well as 3~ML or thicker Co films have an in-plane easy axis, while only 2~ML thick films are magnetized in the out-of plane direction. The first transition is associated with a structural transformation from laterally strained, pseudomorphic 1~ML thick films to relaxed 2~ML thick films.
Our first principles calculations show that the in-plane easy-axis of one- and three-monolayer films is stable with respect to variations of the strain conditions. Only for two-monolayer films, the combination of strain with additional interface and surface effects drives the magnetic easy-axis into the out-of-plane direction.

\begin{acknowledgments} This research was partly supported by the U. S. Department of Energy under contract DE-AC02-05CH11231, by
the Spanish Ministry of Science and Technology under Projects
No.~MAT2003-08627-C02-02, MAT2003-04278 and 2004-HU0010, and by the Comunidad Aut\'{o}noma de Madrid under Project GR/MAT/0155/2004. Additional support by the Hungarian Scientific Research Fund 
(OTKA T037856 and T046267) and the Austrian Ministry of Labor and Economy (bw:aw 98.366) is acknowledged.
\end{acknowledgments}

\end{document}